\pdfoutput=1 

\documentclass[
    ,final            
  ]
  {aipproc}

\layoutstyle{6x9}

\usepackage{graphicx}
\usepackage{caption}

\begin{document}

\title{High energy blazars spectroscopy with X-Shooter on the VLT}


\author{S.~Pita}{address={APC, Univ. Paris Diderot, CNRS/IN2P3,CEA/Irfu, Obs. de Paris, Sorbonne Paris Cit\'e},}
\author{P.~Goldoni}{address={APC, Univ. Paris Diderot, CNRS/IN2P3,CEA/Irfu, Obs. de Paris, Sorbonne Paris Cit\'e},}
\author{C.~Boisson}{address={LUTH, Obs. de Paris, CNRS, Universit\'e Paris Diderot}}
\author{Y.~Becherini}{address={APC, Univ. Paris Diderot, CNRS/IN2P3,CEA/Irfu, Obs. de Paris, Sorbonne Paris Cit\'e},
altaddress={LLR, \'Ecole Polytechnique, F-91128, Palaiseau France}}
\author{L.~G\'erard}{address={APC, Univ. Paris Diderot, CNRS/IN2P3,CEA/Irfu, Obs. de Paris, Sorbonne Paris Cit\'e}}
\author{J.-P.~Lenain}{address={Landessternwarte, Universit\"at Heidelberg, K\"onigstuhl, D-69117 Heidelberg, Germany}}
\author{M.~Punch}{address={APC, Univ. Paris Diderot, CNRS/IN2P3,CEA/Irfu, Obs. de Paris, Sorbonne Paris Cit\'e},}

\begin{abstract}
We present results of observations in the UV to near-IR range for eight blazars, three of which have been recently discovered at Very High Energies (VHE) and five appearing as 
interesting candidates for VHE $\gamma$-ray detection. We focus in this paper on the search for their redshifts, which are unknown or considered as uncertain. 
\end{abstract}

\maketitle

\section{Introduction}
The redshift determination for blazars is often difficult as the bright non-thermal emission of the jet easily hides the host galaxy.
For blazars detected in the VHE range, this measurement is very important as it allows to model the imprint of the Extragalactic
Background Light (EBL) on their emission constraining models of both.  
 
\section{Observations and data reduction}
We briefly present the results of observations\footnote{Based on data collected at the European Southern Observatory under programs P086.B-0.135(A) in 2010 and P088.B-0485(A) in 2011.} 
of eight sources with the X-Shooter spectrograph \cite{verne_2011}, 
which is a single-object medium resolution ($\lambda$/$\Delta$$\lambda$ = 5000 -- 10000) \'echelle spectrograph operating on the VLT. 
The main characteristic of X-Shooter is its unprecedented simultaneous wavelength range from 300 nm to 2400 nm, obtained by splitting the light in three arms: UVB ($\lambda$ = 0.3 -- 0.56 $\mu$m), 
VIS ($\lambda$ = 0.55 -- 1.02 $\mu$m), and NIR ($\lambda$ = 1.00 -- 2.40 $\mu$m). For each source, the observations consist of a set of different exposures of 600 seconds each. 
We processed the spectra using version 1.3 of the X-Shooter data reduction pipeline \cite{goldo_2006}. 
The reduction involves usual steps, including bias subtraction, flat field division with background estimation and wavelength calibration, all performed using day-time calibration frames. 

\section{Results}

\begin{figure}
 \includegraphics[width=0.9\textwidth]{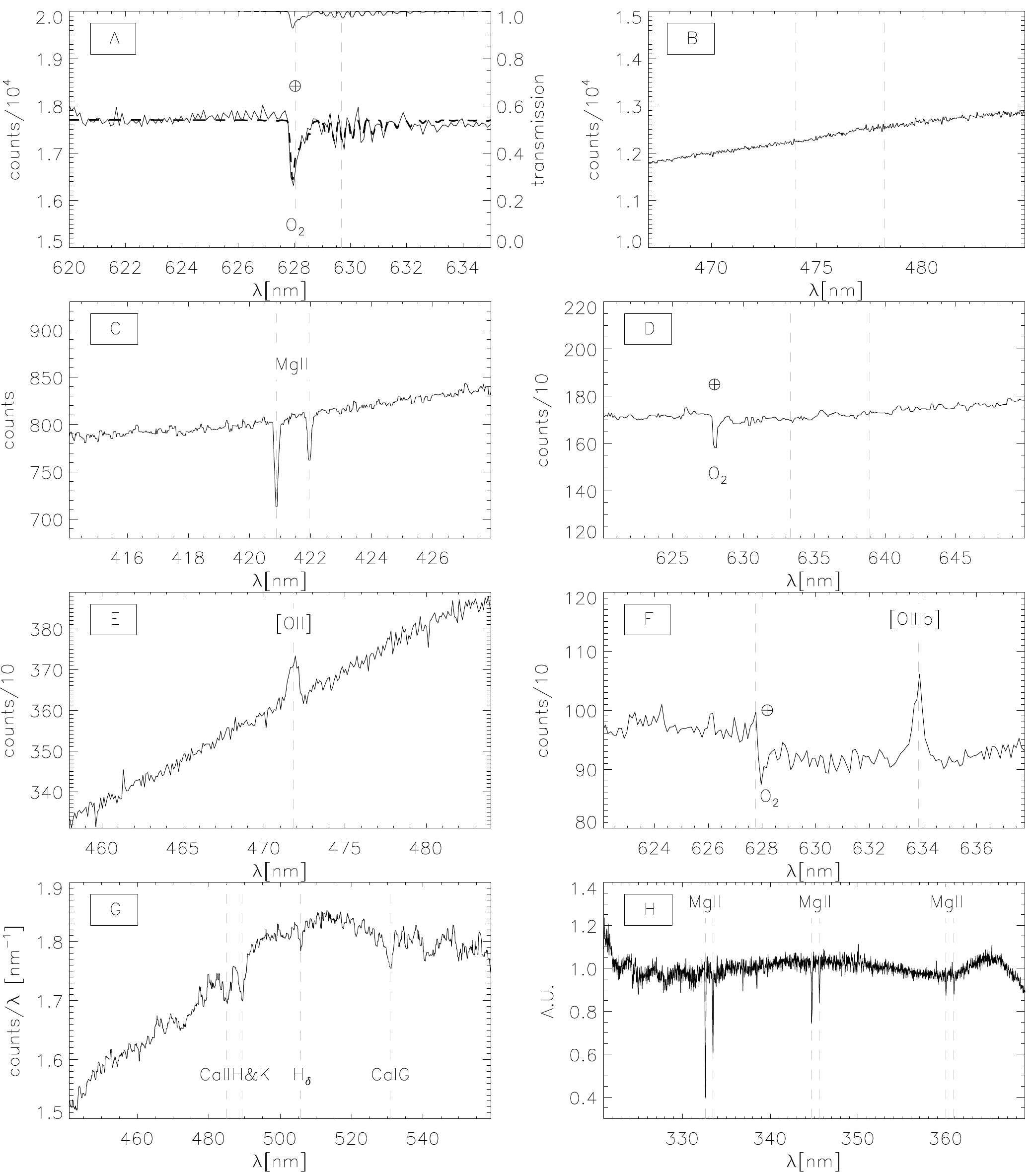}
  \caption{
Selected regions where features (indicated by dashed vertical lines)  were searched for and/or identified for five VHE blazars:
PKS~0447--439 (A \& B), KUV~00311--1938 (C \& D), PKS~0301--243 (E \& F), BZB~J0238--3116 (G) and BZB~J0816--1311 (H). See text for explanations.
}
\label{features}
\end{figure}

{\bf PKS~0447--439}: 
this is one of the brightest blazars dectected by Fermi-LAT \cite{Abdo_2009} and was recently discovered as a VHE emitter by H.E.S.S. \cite{zech_2011}.
Its redshift is still uncertain, with a proposed value z=0.205 \cite{perlman_1998} based on a very weak spectral feature identified as the CaII doublet,
consistent with different non-spectroscopic estimations (\cite{landt_2008}, \cite{zech_2011}, \cite{prandini_2012}).
Conversely, \cite{landt_2012} put a very high lower limit of z>1.246 on its redshift\footnote{
If confirmed, this surprising result would imply that our understanding of VHE $\gamma$-ray propagation is incorrect, 
or that the EBL density is far below the value obtained from galaxy counts.
}, based on weak absorption lines at 6280 \AA\ interpreted as the redshifted MgII 2800 \AA\ doublet. 

The spectrum obtained in this work is strongly dominated by the emission of the nucleus and, despite 
favourable resolution and signal-to-noise levels, does not show any extragalactic spectral line.
The very good signal-to-noise level allows the examination of the weak absorption feature present around 6280 \AA\ (black continuous line in Fig. \ref{features}A), where atmospheric 
molecular absorption is expected from $\mathrm{O_2}$ (indeed, this feature is seen in all our sources, see for example Figs. \ref{features}D and \ref{features}F). 
The expected transmission rate -- smeared to the resolution of our data -- is shown in the top of the same figure.
We see that the observed feature and the expected absorption by $\mathrm{O_2}$, once normalized to the flux level of the source (dashed line), match 
very well, both in shape and wavelength position. This clearly demonstrates that the association of the 6280 \AA\ feature with the Mg II doublet by \cite{landt_2012} 
is not correct\footnote{
In a measurement obtained after this conference, a similar conclusion has been proposed by \cite{fumagalli_2012}.}.
Besides, no absorption is seen where the CaII doublet is expected for a redshift of 0.205 as proposed by \cite{perlman_1998}, see Fig. \ref{features}B, 
although it should be noted that this region is dominated by the non-thermal emission of the nucleus.

{\bf KUV~00311--1938}: identified as a BL Lac by \cite{bauer_2000} and recently detected by H.E.S.S. \cite{Bech_2012}, this source is considered as probably the farthest  
BL Lac detected at VHE, based on a tentative redshift of $z$=0.61 proposed by \cite{pirano_2007}.
From a careful examination of our spectrum, we do not confirm this tentative redshift: as shown in Fig. \ref{features}D, there is no indication of CaII absorption 
at $z$=0.61 (see dashed vertical lines). However, a MgII doublet of narrow lines is clearly identified in the UVB arm around 4215 \AA\, which corresponds to a redshift 
of $z$=0.506 (see Fig. \ref{features}C). This is the only extragalactic feature in our spectrum, it could be due to an intervening system unrelated to the blazar or to
the blazar itself, giving a clear lower limit to the redshift of KUV~00311--1938: $z\ge$0.506.

{\bf PKS~0301--243}: this source has first been identified as a blazar by \cite{impey_1988}, later discovered at high energies by Fermi-LAT \cite{Abdo_2009}, and recently by H.E.S.S. \cite{wout_2012}. 
The best determination of its redshift \cite{falomo_2000} was based on the plausible identification of a single weak emission line with [OIII] 5007 \AA.
We confirm this result and propose a more precise determination of the redshift at $z$=0.266, given the clear detection of the emission lines [OII] 3727.3 \AA\ in the UVB arm 
and [OIII] 5007 \AA\ in the VIS arm (see Figs. \ref{features}E and \ref{features}F, where the dashed lines show the expected positions of these emission lines for $z=$0.266). 
Note that due to an overlap with the atmospheric $\mathrm{O_2}$ absorption feature, one of the [OIII] doublet lines is not clearly detected.

{\bf VHE blazar candidates}: we studied also the spectra of five other blazars, which may be considered as good candidates for TeV emission based on their favourable levels of flux density at 
1.4 GHz ($F_{R}$), their X-ray flux between 0.1 and 2.4 keV ($F_{X}$) and/or the hardness of their spectral slope $\mathrm{\Gamma_{Fermi}}$ as seen by Fermi-LAT. The corresponding numbers and the 
redshift search results are summarized in Table \ref{ztab}. As an illustration, the identified lines for BZB~J0238--3116 and BZB~J0816--1311 are shown in Figs. \ref{features}G 
and \ref{features}H, respectively.
In the case of RBS~334, only a tentative redshift is proposed, as the identified lines are found at the overlap between two arms (UVB and VIS), where we are more subject to systematic errors.

\begin{table}[h]
\begin{tabular}{llllll}
\hline
Source         &  $F_{R}$ & $F_{X}$ & $\Gamma_{Fermi}$ & $z$ & Comments\\
\hline
PKS~0447--439   &  0.73 & 0.025 & 1.86 & -- & $z\geq$1.26 invalidated, 0.205 not confirmed \\
KUV~00311--1938 &  0.04 & 0.026 & 1.76 & $\geq$0.506 & $z=$0.61 not confirmed\\
PKS~0301--243   &  1.43 & 0.018 & 1.94 & 0.266 & [OII] and  [OIII] emission lines \\
\hline
BZB~J0238--3116 &  0.15 & 0.016 & 1.85 & 0.233 & CaII doublet, $H_{\delta}$ and CaIG absorption lines \\
BZB~J0543--5532 &  0.08 & 0.028 & 1.74 & 0.273 & CaII doublet and NaID absorption lines\\
BZB~J0505+0415 &  0.34 & 0.009 & 2.15 & 0.424 & Mg, CaFe and NaID absorption lines\\
BZB~J0816--1311 &  0.14 & 0.027 & 1.80 & $\ge$ 0.288 & Three MgII doublets at different redshifts\\
RBS~334        &  0.05 & 0.007 & 1.56 & 0.411(?) & Indication of the CaII doublet\\
\hline
\end{tabular}
\caption{Radio and X-ray fluxes (relative to corresponding values for PKS~2155--304) and Fermi-LAT spectral slope as reported in BZCat. The redshifts are from this work. See text for details.}
\label{ztab}
\end{table}

\section{Conclusions}
We demonstrate that the lower limit proposed for PKS~0447--439 at $z\ge$1.246 by \cite{landt_2012} is not correct, and find no evidence of the feature used by \cite{perlman_1998} to propose $z=$0.205.
We place a clear lower limit at z$\geq$0.506 to the redshift of KUV~00311--1938, but do not see evidence for the feature used by \cite{pirano_2007} to propose the tentative redshift $z=$0.61.
We confirm the redshift proposed by \cite{falomo_2000} for PKS~0301--243 and improve its precision at $z=$0.266.
Finally, for a set of five blazars considered as good candidates for TeV emission, we give three firmly determined redshifts, a tentative redshift and a lower limit.
We encourage their observation by current VHE $\gamma$-ray observatories.

\begin{theacknowledgments}
S. Pita wishes to thank the ESO staff and in particular C. Martayan for their help in performing the observations.
\end{theacknowledgments}



\bibliographystyle{aipproc}   

\bibliography{xshposter}

\IfFileExists{\jobname.bbl}{}
 {\typeout{}
  \typeout{******************************************}
  \typeout{** Please run "bibtex \jobname" to optain}
  \typeout{** the bibliography and then re-run LaTeX}
  \typeout{** twice to fix the references!}
  \typeout{******************************************}
  \typeout{}
 }

\end{document}